\newcommand{\CLASSINPUTbottomtextmargin}{2cm}
\newcommand{\CLASSINPUTtoptextmargin}{2cm}
\documentclass[conference, a4paper]{IEEEtran_scriptbib}
\IEEEoverridecommandlockouts
\newcommand{\subparagraph}{}
\usepackage[compact]{titlesec}
\titlespacing*{\section}{0.2pt}{0.3\baselineskip}{0pt}
\titlespacing*{\subsection}{0.2pt}{0.0\baselineskip}{0pt}
\usepackage{pgf}
\usepackage{mathtools}
\usepackage{amsmath}
\usepackage{amssymb}
\usepackage{algorithmic}
\usepackage{textgreek}
\usepackage{tikz}
\usepackage{circuitikz}
\usepackage{url}
\DeclareMathOperator{\rect}{rect}
\DeclareMathOperator{\sinc}{sinc}
\DeclareMathOperator{\area}{area}
\newcommand{\bB}{\mathbf{B}}
\newcommand{\bA}{\mathbf{A}}
\newcommand{\bK}{\mathbf{K}}
\newcommand{\bD}{\mathbf{D}}
\newcommand{\sn}{^{(n)}}

\begin{document}
\title{Modeling Shift-Variant X-Ray Focal Spot Blur for High-Resolution
Flat-Panel Cone-Beam CT\vspace{-0.2in}}
\author{\IEEEauthorblockN{Steven Tilley II, Wojciech Zbijewski, Jeffrey H\@. Siewerdsen, J\@. Webster Stayman}
\IEEEauthorblockA{Department of Biomedical Engineering, Johns Hopkins
University, Baltimore, MD\@. Email: web.stayman@jhu.edu}\vspace{-5.0in}
\thanks{This work was supported by NIH grants R21EB014964, R01EB018896,
and an academic-industry partnership with Varian Medical Systems (Palo Alto,
CA). The authors would also like to thank Matthew Jacobson for his assistance in
developing the reconstruction algorithm.}
}
\maketitle

\bstctlcite{IEEEtran:BSTcontrol}

\begin{abstract}

Flat-panel cone-beam CT (CBCT) has been applied clinically in a number of
high-resolution applications. Increasing geometric magnification can potentially
improve resolution, but also increases blur due to an extended x-ray focal-spot.
We present a shift-variant focal-spot blur model and incorporate it into a
model-based iterative-reconstruction algorithm. We apply this algorithm to
simulation and CBCT test-bench data. In a trabecular bone simulation study, we
find traditional reconstruction approaches without a blur model exhibit
shift-variant resolution properties that depend greatly on the acquisition
protocol (e.g.\ short vs.\ full scans) and the anode angles of the rays used to
reconstruct a particular region. For physical CBCT experiments focal spot blur
was characterized and a spatial resolution phantom was scanned and
reconstructed. In both experiments image quality using the shift-variant model
was significantly improved over approaches that modeled no blur or only a
shift-invariant blur, suggesting a potential means to overcome traditional CBCT
spatial resolution and system design limitations.

\end{abstract}

\section{Introduction}

\looseness=-1
Flat-panel cone-beam CT (CBCT) is a promising modality for high-resolution
applications, such as quantitative trabecular bone analysis in extremities
imaging and microcalcification detection in
mammography. Current application-specific imaging systems are often
unable to resolve all trabeculae or microcalcifications, which can be on the
order of 100\textmu m. A high-magnification
geometry has the potential to improve resolution, but projections suffer from increased
blur due to the extended focal spot. Model-based iterative reconstruction
(MBIR) methods have previously demonstrated improved image quality through the use
of sophisticated system and noise models.
Proper modeling of the x-ray focal spot, and
incorporation of this model into a MBIR method, can mitigate the effects of focal
spot blur in high-resolution high-magnification reconstructions.

\looseness=-1
Previously, we have developed a reconstruction method that models detector blur,
focal spot blur, and spatial noise correlations using a staged approach (deblurring and
other preprocessing followed by reconstruction).\cite{TilleyII2015a}
Shift-invariant blur models were assumed in order to simplify deblurring. However, such
assumptions are not valid at large fan
angles, where the angulation of the anode results in a position-dependent
apparent focal-spot shape. Moreover, this effect is more pronounced in
high-magnification systems due to a larger focal-spot blur. Properly
modeling shift-variant focal-spot blur is critical to generating high-resolution
images in these systems. Previous work by La Rivi\`{e}re to model shift-variant
focal-spot blur addressed deblurring data for multidetector CT systems with the
anode-cathode axis of the x-ray source oriented axially.\cite{LaRiviere2007}

In
this work, we characterize focal spot blur along the anode-cathode axis in a
CBCT system where this axis is
perpendicular to the axis of rotation (a common orientation in CBCT systems). We use a non-linear objective function
that includes shift-variant blur in the forward model (e.g. no deblurring in preprocessing)
to reconstruct high-resolution objects in simulation and
test-bench studies.

\section{Methods}

\subsection{Forward Model and Objective Function}

We use the general forward model:
\begin{equation}
    y \sim \mathcal{N}(\bB\exp(-\bA\mu), \bK_Y)
    \label{eq:forwardmodel}
\end{equation}
with measurement vector, $y$, and object attenuation values, $\mu$.
The linear operator $\bB$ contains focal spot blur
and gain terms (e.g. photons per pixel), $\bA$ is the forward projector, and $\bK_Y$ is
the measurement covariance matrix. The corresponding penalized-likelihood objective function
is:
\begin{equation}
    \hat{\mu} = \arg \min ||y - \bB \exp(-\bA \mu)||_{\bK_Y^{-1}}^2 + \beta
    R(\mu)
    \label{eq:objective}
\end{equation}
where $R$ is a penalty function and $\beta$ is the penalty strength. 

Equation~\eqref{eq:objective} was minimized using a separable paraboloid
surrogates approach, similar to that of Erdo{\u g}an et
al.\cite{Erdogan1999,erdogan1999a} but with an added separability step in
the $\bB\exp(-\bA\mu)$ term. The resulting baseline algorithm is:
\begin{algorithmic}
\STATE $a = \bB^T\bK_Y^{-1}\bB1$, $\gamma = \bA1$, $b = \bB\bK_Y^{-1}y$
\FOR{$n = 1:N$}
    \STATE{$l\sn = \bA \mu\sn$}
    \STATE{$d\sn = -b - \bD\{a\}\exp(-l\sn) + \bB^T \bK_Y^{-1} \bB\exp(-l\sn)$}
    \STATE{$h\sn_j(l_j) \triangleq 0.5 a_j \exp(-2l_j) + \exp(-l_j)d_j\sn$}
    \STATE $c_j\sn = $ optimum curvature of $h\sn_j$ from \cite{Erdogan1999}
    \STATE{$L\sn = \bA^T(-\bD\{a\} \exp(-2 l\sn) - \bD\{d_n\} \exp(- l\sn))$}
    \STATE{$c_\mu\sn = \bA^T \bD\{\gamma\}c\sn$}
    \STATE{$\mu^{(n + 1)} = \left[\mu\sn + \frac{-L\sn - \beta \bigtriangledown R|_{\mu\sn}}{c\sn_\mu +
    \beta \bigtriangledown^2 R|_{\mu\sn}} \right]_+$}
\ENDFOR
\end{algorithmic}

We further extend the algorithm using Nesterov's acceleration method.
All reconstructions used 20 ordered subsets.\cite{erdogan1999a} The
regularization gradient and curvature were computed using standard surrogate
techniques.\cite{erdogan1999a}

\subsection{Shift-Variant Blur Model}

\looseness=-1
We model the shift-variant focal spot blur along directions parallel to the
detector. The model approximates a depth-independent blur. (See
\S~\ref{sec:discussion} for a discussion of depth-dependent effects.) Therefore, the blur model
can be included in the $\bB$ term in \eqref{eq:forwardmodel}.
To estimate a continuous source-blur model for discrete inputs and outputs, we
use nearest neighbor interpolation to create a continous approximation of the
input image, apply a convolution operation, then discretize the signal using a
rectangular kernel with the dimensions of a pixel and sampling at pixel centers.
The full operation is:
\begin{multline}
    g[k, l] = \int_{x, y} \int_{\xi, \eta} \sum_{i, j} f[i, j] \rect(\frac{\xi -
    i T_x}{T_x}, \frac{\eta - j T_y}{T_y}) T_x T_y\\
    h(x, y; \xi, \eta) \rect(\frac{x - l T_x}{T_x}, \frac{y - k
    T_y}{T_y}) \, d\xi \, d\eta \, dx \, dy
    \label{eq:fullconv}
\end{multline}
where $f$ and $g$ are the input and output images, $T_x$ and $T_y$ are the pixel widths along the corresponding directions
and $h(\cdot, \cdot; \xi, \eta)$ is the impulse response of a point source at
$\xi$, $\eta$.
Equation~\eqref{eq:fullconv} can be approximated by discretizing variables and assuming $h$ is constant
over small displacements. We sample $x$ and $\xi$ at intervals of $T_x / s$ and
$y$ and $\eta$ at intervals of $T_y / s$, where $s$ is an odd integer. Applying
these approximations and simplifying leads to:
\begin{multline}
    g[k, l] \approx \sum_{\mathclap{j, i, a, b}} f[i, j]
    |1 - a||1 - b|\\
    h \left( (a + l) T_x, (b
    + k) T_y;i T_x, j T_y \right) T_x T_y / s^2
    \label{eq:finalblur}
\end{multline}
where $a$ and $b$ range from $-(s - 1) / s$ to $(s - 1) / s$ in increments of
$1/s$.
The transpose operation (e.g. for $\bB^T$) requires switching the indices for $f$ and $g$, and
summing over $k, l$ instead of $i, j$.

\begin{figure}
    \begin{center}
        \begin{circuitikz}
            \def\sa{0.342}
            \def\ca{0.94}
            \draw[line width=3.0](\sa * 0.5, -\ca * 0.5) -- (-\sa * 0.5, \ca * 0.5);
            \node at (1.3, 0.1) {Focal Spot};
            \draw[dashed] (-\sa * 0.5, \ca * 0.5) -- (-\sa * 0.5, \ca * 0.5 - 1);
            \draw[dashed] (\sa * 0.5, -\ca * 0.5) arc (290:270:1);
            \node at (\sa * 0.1, -\ca * 0.75) {$\theta$};
            \draw (-3, -\ca) -- (\sa, -\ca) -- (-\sa, \ca) -- (-3, \ca);
                        \node at (-1.0, 0.0) [text width=1.5cm]{Anode Coords.};
            \draw[very thick] (-3.0, 5.0) -- (3.0, 5.0) node[above, pos=0.5] {Detector};
            \draw (-3.0, 5.4) node[above right] {Anode Side};
            \draw (3.0, 5.4) node[above left] {Cathode Side};
            --cycle;
            \draw (2.5, 2.5) -- (1.1, 2.5);
            \draw (.9, 2.5) -- (-0.5, 2.5);
            \draw[dashed] (\sa * 0.5, -\ca * 0.5) -- ({-(5.0 + 0.5 * \ca) / (2.5 + 0.5 *
            \ca) * (-1.0 + 0.5 * \sa) + 0.5 * \sa}, {(5.0)});
            \draw[dashed] (-\sa * 0.5, \ca * 0.5) -- ({-(5.0 - 0.5 * \ca) / (2.5 - 0.5 *
            \ca) * (-1.0 - 0.5 * \sa) - 0.5 * \sa}, {(5.0)});
            \draw[dashed] (0, 0) -- (2.0, 5.0);
            \draw[dashed] (0, 0) -- (0.0, 5.0);
            \fill (0.0, 5.0) circle[radius=.10] node[below right] {($u_0$,
            $v_0$, $0$)};
            \draw (0.2, 5.0) -- (0.2, 4.8) -- (0, 4.8);
            \draw[very thick, ->] (-2.0, 4.0) -- (-1.0, 4.0) node[right] {$u$};
            \draw[very thick, ->] (-2.0, 4.0) -- (-2.0, 3.0) node[below] {$w$};
            \fill (-1.5, 3.5) circle[radius=0.08];
            \draw (-1.5, 3.5) circle[radius=0.15] node[below right] {$v$};
            \draw (-1.6, 4.05) node[above] {Detector Coordinates};
            \fill (1.0, 2.5) circle[radius=.10] node[below right] {($u_p$,
            $v_p$, $w_p$)};
            \fill (0.0, 0) circle[radius=.10] node[below right] {\hspace{6pt}($u_0$,
            $v_0$, $SDD$)};
            \fill (2.0, 5.0) circle[radius=.10] node[above] {($u_c$, $v_c$, $0$)};
            \draw[very thick, ->] (-3.0, 0.3) -- (-3.0 + \sa, 0.3 - \ca)
            node[right] {$x$};
            \draw[very thick, ->] (-3.0, 0.3) -- (-3.0 + \ca, 0.3 + \sa)
            node[right] {$z$};
            \draw (-2.3, 0.0) circle[radius=0.15] node[below right] {$y$};
            \draw (-2.3 + .15 / 1.4, 0.0 + .15 / 1.4) -- (-2.3 - .15 / 1.4, 0.0 -
            .15 / 1.4);
            \draw (-2.3 - .15 / 1.4, 0.0 + .15 / 1.4) -- (-2.3 + .15 / 1.4, 0.0 -
            .15 / 1.4);
            \draw (-1.6, 1.05) node[above] {Anode};
            \draw (3.0, 1.0) to[L] (3.0, -1.0);
            \node at (3.0, 1.2) {Cathode};
            
        \end{circuitikz}
    \end{center}
    \caption{Geometry used to calculate the focal spot blur impulse
    response. The focal spot is represented by the bold line on the side of the
    anode. All coordinates are in detector coordinates. The origin of the anode
    coordinate system is at ($u_0$, $v_0$, $SDD$).}\label{fig:geo}
\end{figure}
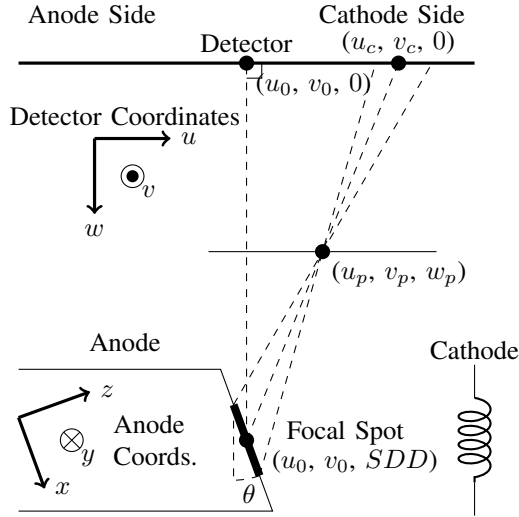

\looseness=-1
The impulse response ($h$) centered at a given point ($u_c$, $v_c$) is assumed to be a binary
function, with values either equal to $0$ or $k = \area(h(\cdot, \cdot; u_c,
v_c))^{-1}$. To determine the value of $h(u, v; u_c, v_c)$, the point ($u$, $v$)
is backprojected through a pinhole onto the anode. A two dimensional cross section of the
geometry is illustrated in Figure~\ref{fig:geo}. The pinhole is placed a distance
$w_p$ from the detector and along the line connecting ($u_c$, $v_c$) with the
center of the focal spot. If the backprojected point is in the rectangular focal
spot, $h(u, v; u_c, v_c) = k$, otherwise $h(u, v; u_c, v_c) = 0$. The area of $h$ was found by forward
projecting the corners of the focal spot through the pinhole, and applying
Bretschneider's formula to the resulting points.\cite{weisstein_bretschneiders}

\subsection{Simulation Study}

\begin{figure}
    \begin{center}
        \input{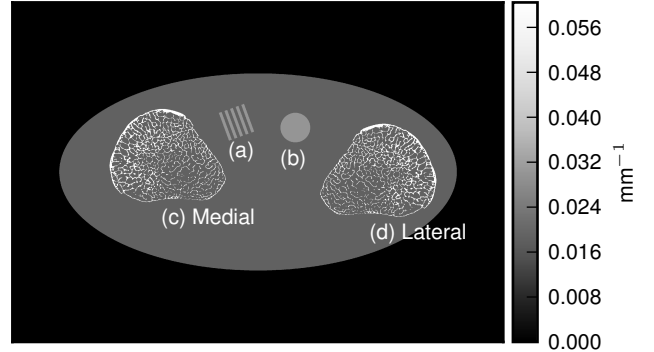}
        \caption{Digital extremeties phantom with medial (c) and lateral (d)
        bones, line pairs (a), and a uniform
        disc (b).}\label{fig:phantom}
    \end{center}
\end{figure}

\looseness=-1
Data were generated from the digital extremities phantom in
Figure~\ref{fig:phantom}. Line integrals were generated from a high-resolution
truth image (3300$\times$2300 image of 30\textmu m voxels) projected onto a
one-dimensional detector with 8192 pixels and a 48.5\textmu m pixel pitch. A 
high-magnification geometry was used, with a source-detector distance of 1200mm, a
source-axis distance of 250mm, and an angular spacing of 0.5$^\circ$. These
line integrals were downsampled by a factor of 4 to give a pixel pitch of
194\textmu m. Measurements were generated from the downsampled line integrals
($l$) according to:
\begin{align}
        &y_{\text{noiseless}} = \bB_s I_0 \exp(-l)\\
        &y = y_{\text{noiseless}} + \mathcal{N}(0, \bD\{y_{\text{noiseless}}\})
        + \mathcal{N}(0, \bD\{\sigma_{ro}^2\})
\end{align}  where $I_0$ is $10^4$ photons per pixel, $\bB_s$ is the focal
spot blur operator (we assume there is no detector blur), and the readout-noise standard deviation ($\sigma_{ro}$) is 3.32 photons.
The focal spot was modeled as a
5mm$\times$0.8mm rectangle on a 14$^\circ$ anode with the anode-cathode axis
parallel to the detector row. The sampling factor ($s$) was equal to 41.
(Note that $\bB_s I_0$ is equivalent to $\bB$ in \eqref{eq:forwardmodel}.)
Data were generated using two short scans (short-1 and short-2) spaced 180
degrees apart, and a full scan. The
short-1 scan placed the medial bone (Figure~\ref{fig:phantom}c) predominately
on the anode side, and the lateral bone
(\ref{fig:phantom}d) predominately on the cathode side. The
reverse is true for the short-2 scan.

Data from each scan were reconstructed using
the algorithm presented above with three models for focal-spot blur: identity
(no blur), shift-invariant blur, and shift-variant blur.
The sampling factor ($s$) used in reconstructions was
11. Data were reconstructed into a 1650$\times$1150 volume of 60\textmu m voxels
using a Huber penalty ($\delta = 10^{-4}$). The covariance matrix was modeled as
$\bD\{y + \sigma_{ro}^2\}$.

The accuracy of trabeculae segmentation in the medial and lateral bones was used
as a measure of image quality. The truth segmentation for each bone was found by
downsampling the high-resolution phantom to match the reconstruction volume
dimensions and thresholding at the average attenuation of bone and fat. Data
were reconstructed at several penalty strengths and thresholded at regularly
spaced values between the attenuation values of fat and bone. Accuracy was
quantified as the mutual overlap between the thresholded truth, $t$, and the thresholded
reconstruction, $r$:\cite{sonka_ise_2007}
\begin{equation}
    \text{mutual overlap}(t, r) = 2(\sum t r)(\sum(t + r))^{-1}
\end{equation}

\subsection{Bench Characterization}\label{sec:methods:mtf}

\looseness=-1
To apply the approach to physical data, we characterized the focal spot blur on a
CBCT test bench consisting of a Rad-94
x-ray tube (Varian, Salt Lake City UT), a PaxScan 4343CB flat-panel detector (Varian,
Palo Alto CA), and a SDD of 108 cm.
In this work we focus on two-dimensional reconstructions, and therefore only
measure one dimensional MTFs 
along the $u$ axis.
MTFs were measured using a tungsten edge\cite{TilleyII2015a}\cite{Samei1998}
placed at isocenter (40cm from the source) and translated in the $\pm u$ directions.
The detector MTF was measured by placing the edge at the  detector. We assume the detector MTF is shift-invariant and fit it to the following
model:\cite{siewerdsen_signal_1998}
\begin{equation}
        |MTF_d(f_u)| = \left| \frac{\sinc(f_u
        T_x)}{1 + H f_u^2} \right|
\end{equation}
where $f_u$ is the spatial frequency in
mm$^{-1}$ and $H$ is a blur parameter. The focal spot MTF at each position $u_p$ was modeled as a rect function with an apparent
length $L(u_p)$, resulting in the combined MTF:
\begin{equation}
        |MTF_{sd}|(f_u; u_p) = \left| \sinc(f_u L(u_p)) MTF_d(f_u) \right|
\end{equation}
Theoretical apparent blur lengths from anode angle
($\theta$) and focal spot length ($L$) were fit to
the measured lengths to yield estimates for $\theta$ and $L$.

\subsection{Resolution Phantom Study}

\looseness=-1
A cylindrical resolution phantom (CatPhan CTP528 High Resolution Module, Phantom Laboratory, Salem, NY) 
with variable frequency line pairs 
was scanned on the CBCT test bench. The source-detector and source-axis
distances were 108 cm and 40 cm respectively. A full scan of 720 projections was
collected at 80 kVp and 0.504 mAs per projection. Data were reconstructed using
the identity and shift-variant blur models, as well as three shift-invariant blur
models. The three shift-invariant blurs modeled were the blur at the center of
the detector (as in the simulation study) and the blur at either edge of the
detector. The presented MBIR algorithm was used with 800 iterations to ensure a
nearly converged solution.
The reconstruction volume
was 170mm$\times$170mm with 100\textmu m voxels. The blur model used the focal spot
length and anode angle from \S~\ref{sec:methods:mtf} and a subset parameter
($s$) of 5. We assume that detector blur is negligible and do not model it in the
reconstruction algorithm.

\section{Results}

\subsection{Simulation Study}

\begin{figure}
    \input{betasegqual.pgf}
    \caption{Best mutual overlap versus $\beta$. A) Medial and B)
    Lateral bone.}\label{fig:betasegqual}
\end{figure}

The best mutual overlap values (over all threshold values) for each ($\beta$)
are shown in Figure~\ref{fig:betasegqual}. Results are shown for reconstructions
with an identity (ID) blur model and the shift-variant (SV) blur
model. Each line represents a blur model and scan type combination,
and each point represents a reconstruction. A higher best mutual overlap
indicates that a segmentation based on that reconstruction is closer to the
truth segmentation, and the reconstruction is therefore more accurate. All
methods that used the SV model were more accurate than those that
used the ID model, which is evident by comparing the maximum of each
line. With the ID model, the best quality
segmentation of the medial bone is achieved with data from the short-1 scan,
which placed the medial bone projections primarily on the high-resolution
(anode) side
of the detector. The lowest quality was
the short-2 scan, which placed the projections primarily on the low-resolution
(cathode)
side. The full scan reconstructions with the ID model
rank between the reconstructions from the two short scans. Neglecting to model
blur is equivalent to assuming that classically redundant projections in the
full scan (i.e\@. those with the same integration path but reversed direction)
contain the same information, despite the fact that they are subject to
different degrees of blurring, which results in a reconstruction whose
image quality is a compromise between that of the two short scan
reconstructions. Predictably, the lateral bone reconstructions are best when
using the short-2 scan and worst when using the short-1 scan, in which the
lateral bone projection data was on the high- and low-resolution sides of the
detector, respectively.

\looseness=-1
When using the SV model, the full scan provides the best
reconstruction of both bones, followed by the short-1 scan and then the short-2 scan
in the case of the medial bone, and the short-2 scan and then the short-1 scan for
the lateral bone. The better image quality of the full scan images over the
corresponding high-resolution short-scan reconstructions can be attributed to the additional
(low-resolution) data. The SV model can use this additional
information to improve the reconstruction without losing details provided by the
high-resolution data. In effect, rather than averaging the redundant data, the low-frequency data
is used to reduce noise while the high-frequency data maintains spatial resolution.
The corresponding low-resolution scan for each bone (short-2
for the medial bone and short-1 for the lateral bone) results in the lowest
quality reconstructions due to the increased difficulty in deblurring the data.

\begin{figure}
    \input{simrecons.pgf}
    \caption{Reconstructions of the medial bone with the highest mutual overlap over all thresholds
    and $\beta$'s. The top half of each reconstruction is thresholded.}\label{fig:simrecons}
\end{figure}

\looseness=-1
Figure~\ref{fig:simrecons} shows the medial-bone reconstructions (bottom of each image) and
segmentations (top of each image) corresponding to the best possible mutual overlap
(optimal threshold and $\beta$ values) with each scan type and blur model
combination. All SV reconstructions depict more
trabecular structure than
the shift-invariant (SI) or ID models. The difference in image quality among ID reconstructions is
readily apparent in these images, with the short-1 scan resulting in the most
trabecular detail. Finally, the SI images
depict more detail than the ID model but less detail than
the SV reconstructions. However, the SI model
results in a ringing artifact, particularly evident on the lower left aspect
of the medial bone in the full scan reconstruction. This is likely due to
blur/model mismatch (the SI model is accurate at the
center of the detector but less accurate at the edges).

\subsection{Focal Spot Measurement}

\begin{figure}
    \input{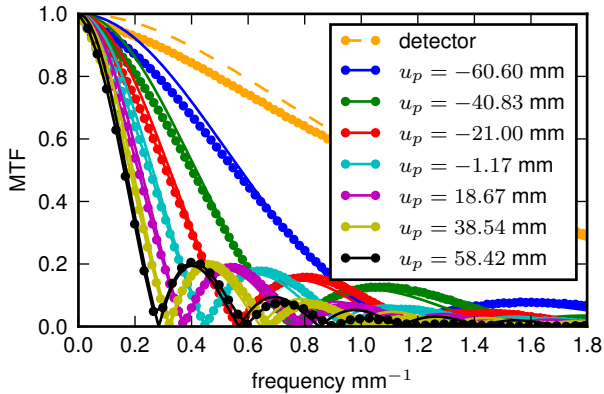}
    \caption{MTFs and fits for the detector and the detector+source blur at different
    displacements from the center of rotation.}\label{fig:mtfs}
\end{figure}

\looseness=-1
The detector MTF and the combined focal-spot and detector MTFs at different
positions are shown in Figure~\ref{fig:mtfs}. The magnification in this system
was about 2.7, so that the focal-spot blur dominates over the
detector blur. Each combined focal-spot and detector MTF is labeled by the
distance of the tungsten edge from the central ray. At positive
positions, the edge is on the cathode side, and at negative positions the edge
is on the anode side. There is a dramatic difference in MTFs at different
positions due to the angulation of the anode. Fits for each MTF are also shown.
These fits give the apparent length of the focal spot at each position, which was
used to estimate the actual length of the focal spot and the angle of the anode. The
focal spot length was found to be 5.23mm and the anode angle was 14.3$^\circ$.

\subsection{Bench Study}

\begin{figure}
    \input{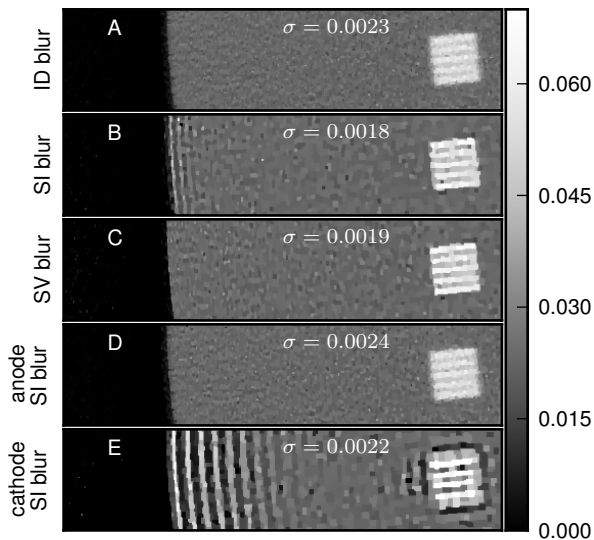}
    \caption{Physical CBCT reconstructions. Each subfigure shows a portion of the phantom
             from the edge to one of the line pairs. Each reconstruction has
             approximately the same noise level (indicated in each subplot in
             units of
             mm$^{-1}$ and denoted by $\sigma$).}\label{fig:realrecs}
\end{figure}

Figure~\ref{fig:realrecs} shows the same region-of-interest of five
reconstructions, each of which used a different blur model. The three
SI blur models are the apparent focal spot size at the center,
anode side, and cathode side of the detector.
The reconstructions have approximately the same amount of noise (estimated by
computing standard deviation in a flat region at the center of the phantom). The
line pairs in the SV (\ref{fig:realrecs}c) and center SI
(\ref{fig:realrecs}b) reconstructions are much
sharper than those in the ID reconstruction (\ref{fig:realrecs}a). That the
SI reconstruction line-pairs are roughly as sharp as those of the SV reconstruction suggests that at this distance from isocenter
(approximately 4.75cm) the SI approximation is fairly accurate.
However, at the edge of the phantom (approximately 7.5cm from isocenter), this
assumption breaks down, and the resulting mismatch between the model and the
actual blur results in a ``ringing'' artifact. The anode-side SI
blur model (\ref{fig:realrecs}d) underestimates the blur over most of the detector, reducing ringing
compared to the center SI blur model but also reducing the
sharpness of the line pairs. The cathode-side SI blur model
(\ref{fig:realrecs}e) overestimates the blur over much of the detector,
increasing the ringing artifact.

\section{Discussion}\label{sec:discussion}

The image quality difference in identity blur model reconstructions from the two
different short scans illustrates the importance of considering shift-variance
in high-resolution, high-magnification systems. The poor image quality and/or
ringing artifact in the
reconstructions with a shift-invariant blur model demonstrate that this model
is a poor approximation for large objects (relative to the field of view), and
that a full shift-variant model is more appropriate. These results also suggest a
means to improve local resolution properties when advanced blur models are not
available: if the location of a high resolution target in the object is known
\textit{a priori,} then that object can be placed such that the high resolution
target favors the anode side of the detector during a short scan.

\looseness=-1
This work suggests x-ray tube orientation is an
important factor in system design. Blur
shift-variance, and therefore reconstruction resolution, will depend on whether
the anode-cathode axis is oriented parallel or perpendicular to the axis of
rotation. Models such as the one presented may alter the trade-off associated
with tube orientation, allowing for more flexibility in system design.
Future studies
will analyze three-dimensional reconstructions in order to properly
characterize resolution/image quality both in-plane and axially. While we have demonstrated
the utility of a depth-independent source blur model, future work
will consider depth-dependent source blur effects. In the presented bench study, we estimate that
apparent focal spot size approximately doubled over the length of the object along the
source-detector direction. By comparison, the measured
apparent focal-spot lengths approximately quadrupled over the length the object
along the direction parallel to the detector. Thus, depth-dependent shift-variance is a large
effect, but not as large as shift-variance due to anode angulation.

We have provided a method to improve image quality with an
advanced shift-variant blur model, and used this model to reconstruct high-resolution 
trabecular details in a simulation study and fine
line-pair patterns on a CBCT test bench. This technique could help overcome
spatial resolution limits in high-magnification systems, improving
current systems and allowing new systems to be designed with higher
magnifications for high-resolution applications.
\bibliographystyle{IEEEtrancompact}
\bibliography{ZoteroLibrarybibtex,IEEEcontrol}

\end{document}